\documentclass[twocolumn]{aastex62}
\usepackage{amsmath}     
\usepackage{wasysym}           
\usepackage{graphicx}
\usepackage{amssymb}
\usepackage{epstopdf}
\usepackage{mathrsfs}
\usepackage{anyfontsize}
\usepackage{natbib}
\usepackage{color}
\usepackage{lipsum}
\usepackage{diagbox}
\usepackage[percent]{overpic}

\shorttitle{Core Reformation}
\shortauthors{Nixon \& Coughlin}
\begin{document}
\title{Stellar Revival and Repeated Flares in Deeply Plunging Tidal Disruption Events}

\author[0000-0002-2137-4146]{C.~J.~Nixon}
\affiliation{Department of Physics and Astronomy, University of Leicester, Leicester, LE1 7RH, UK}

\author[0000-0003-3765-6401]{Eric R.~Coughlin}
\affiliation{Department of Physics, Syracuse University, Syracuse, NY 13210, USA}

\email{cjn@leicester.ac.uk, ecoughli@syr.edu}

\begin{abstract}
Tidal disruption events with tidal radius $r_{\rm t}$ and pericenter distance $r_{\rm p}$ are characterized by the quantity $\beta = r_{\rm t}/r_{\rm p}$, and ``deep encounters'' have $\beta \gg 1$. It has been assumed that there is a critical $\beta \equiv \beta_{\rm c} \sim 1$ that differentiates between partial and full disruption: for $\beta < \beta_{\rm c}$ a fraction of the star survives the tidal interaction with the black hole, while for $\beta > \beta_{\rm c}$ the star is completely destroyed, and hence all deep encounters should be full. Here we show that this assumption is incorrect by providing an example of a $\beta = 16$ encounter between a $\gamma = 5/3$, solar-like polytrope and a $10^6 M_{\odot}$ black hole -- for which previous investigations have found $\beta_{\rm c} \simeq 0.9$ -- that results in the reformation of a stellar core post-disruption that comprises approximately 25\% of the original stellar mass. We propose that the core reforms under self-gravity, which remains important because of the compression of the gas both near pericenter, where the compression occurs out of the orbital plane, and substantially after pericenter, where compression is within the plane. We find that the core forms on a bound orbit about the black hole, and we discuss the corresponding implications of our findings in the context of recently observed, repeating nuclear transients. 
\end{abstract}


\section{Introduction}
\label{sec:intro}
The impact of the tidal gravitational field of one massive body on another can be separated into two distinct regimes: perturbative and destructive. In the former, tidal effects are small and their consequences can be understood in the context of linear perturbation theory: the self-gravity and pressure gradient of the perturbed body are each large (but, by virtue of approximate hydrostatic balance in the presence of the perturber, nearly cancel one another as concerns the force on any given fluid element) relative to the tidal term, and hence the ratio of the magnitude of the tidal gravitational field to the self-gravitational field yields a natural smallness parameter about which to perturb the system. The methods for analyzing this limit are well-established (e.g., \citealt{fabian75, press77, lee86, ogilvie14}).

In the second limit, the tidal effects of one (assumed more massive) body on the other (less massive) object are large relative to the pressure and self-gravity of the latter; from here we restrict our discussion to the scenario in which the less massive object is a star of mass $M_{\star} \sim 1M_{\odot}$ and radius $R_{\star} \sim 1R_{\odot}$ and the more massive object is a supermassive black hole of mass $M_{\bullet} \sim 10^{6}M_{\odot}$. In this case, it is natural to assume that while the star is within the distance to the black hole such that this inequality between tides and self-gravity+pressure holds, one may approximate the motion of the stellar material as ballistic. This approximation has been made by a number of authors, e.g., \citet{Carter:1983aa, Bicknell:1983aa, Stone:2013aa}, in analyzing the ``deeply plunging'' limit of a star destroyed by the tides of a supermassive black hole, known as a tidal disruption event (TDE).

The existence of these two extremes naturally leads to the concept of a ``critical distance'' that separates the survival and destruction of the star, which, upon equating the tidal force of the black hole to the self-gravity of the star, should be on the order of $r_{\rm t} = R_{\star}\left(M_{\bullet}/M_{\star}\right)^{1/3}$. The ratio $r_{\rm t}/r_{\rm p} \equiv \beta$, where $r_{\rm p}$ is the point of closest approach of the star to the black hole, then quantifies the degree to which the star is modified by tides: $\beta \ll 1$ is in the perturbative regime, whereas $\beta \gg 1$ is in the destructive regime. The ``critical $\beta$'' that separates full from partial stellar disruption\footnote{In a partial disruption, the low-density, outer envelope of the star is stripped off while the dense interior remains intact.} as a function of (e.g.) stellar properties has now been investigated by a number of authors (first by \citealt{Guillochon:2013aa}) and is a cornerstone of TDE theory: there is a $\beta$ below (above) which the star does (does not) survive the tidal encounter with the black hole.

Here we show with a specific counterexample that this widely held notion is false. We discuss the results of a smoothed-particle hydrodynamics (SPH) simulation of a $\beta = 16$ encounter -- first presented in \citet{Norman:2021aa} -- between a solar-like star (i.e., a star with a solar mass and radius modeled as a $\gamma = 5/3$ polytrope) and a $10^6 M_{\odot}$ black hole. Despite being well past the previously established limit of full disruption, which, as demonstrated by\footnote{\citet{Mainetti:2017aa} used three different numerical methods and found that the critical $\beta$ for full disruption of a $\gamma = 5/3$ polytrope was between $0.91$ and $0.94$ (see their Table 2), the critical $\beta$ found by \citet{Guillochon:2013aa} was $0.90$, and \citealt*{Miles:2020aa} used {\sc phantom} (the same code used here) to infer a critical $\beta$ of $\sim 0.92$ at both $10^6$ and $10^7$ particles. } \citealt{Guillochon:2013aa}, \citealt{Mainetti:2017aa}, and \citealt{Miles:2020aa}, occurs at $0.90 \lesssim \beta \lesssim 0.94$ for this type of star, a core containing $\gtrsim 22.8\%$ of the mass\footnote{This quantity was measured from the simulation at a time of approximately one week (corresponding to $1.3\times 10^5GM_\bullet/c^3$) after the original star reached pericentre, and in the simulation that employed 128\,M particles to model the star. At this time the core has a maximum density that is just over three orders of magnitude greater than the densest parts of the rest of the debris stream, and we therefore identify particles belonging to the core, and thus contributing to its mass estimate, as those with density greater than 0.1\% of this maximum density. This criterion yields a mass estimate of 22.87\%. For comparison, taking a density cut of 1\% yields a mass estimate of 22.79\%. Because it is still accreting material at this time, its mass will grow above this value by a small amount.} of the original star reforms out of the disrupted debris. 

\section{Results}
\label{sec:simulations}
\begin{figure*}
	\centering
	\includegraphics[width=0.49\textwidth]{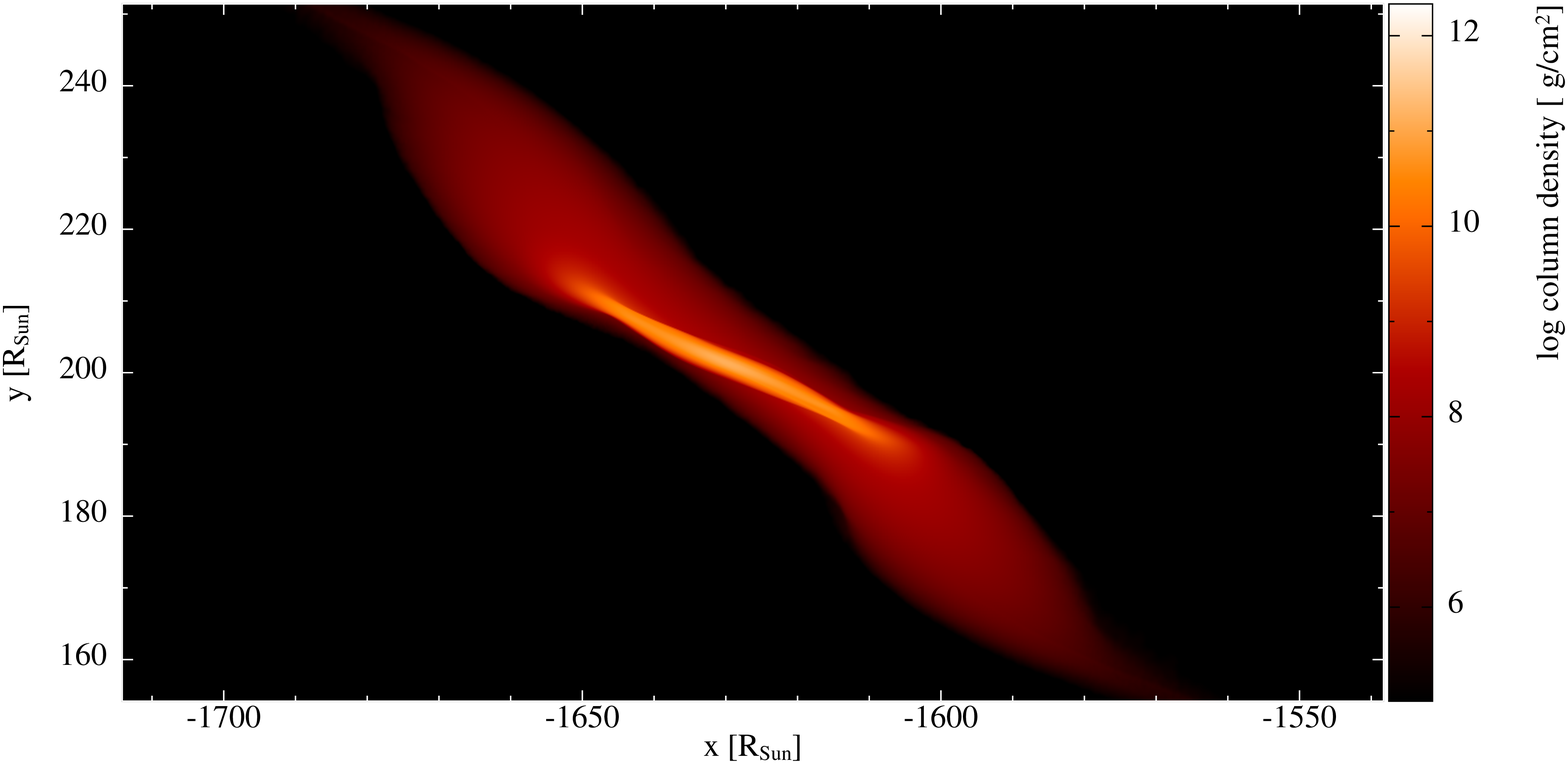}\hfill
	\begin{overpic}[width=0.49\textwidth]{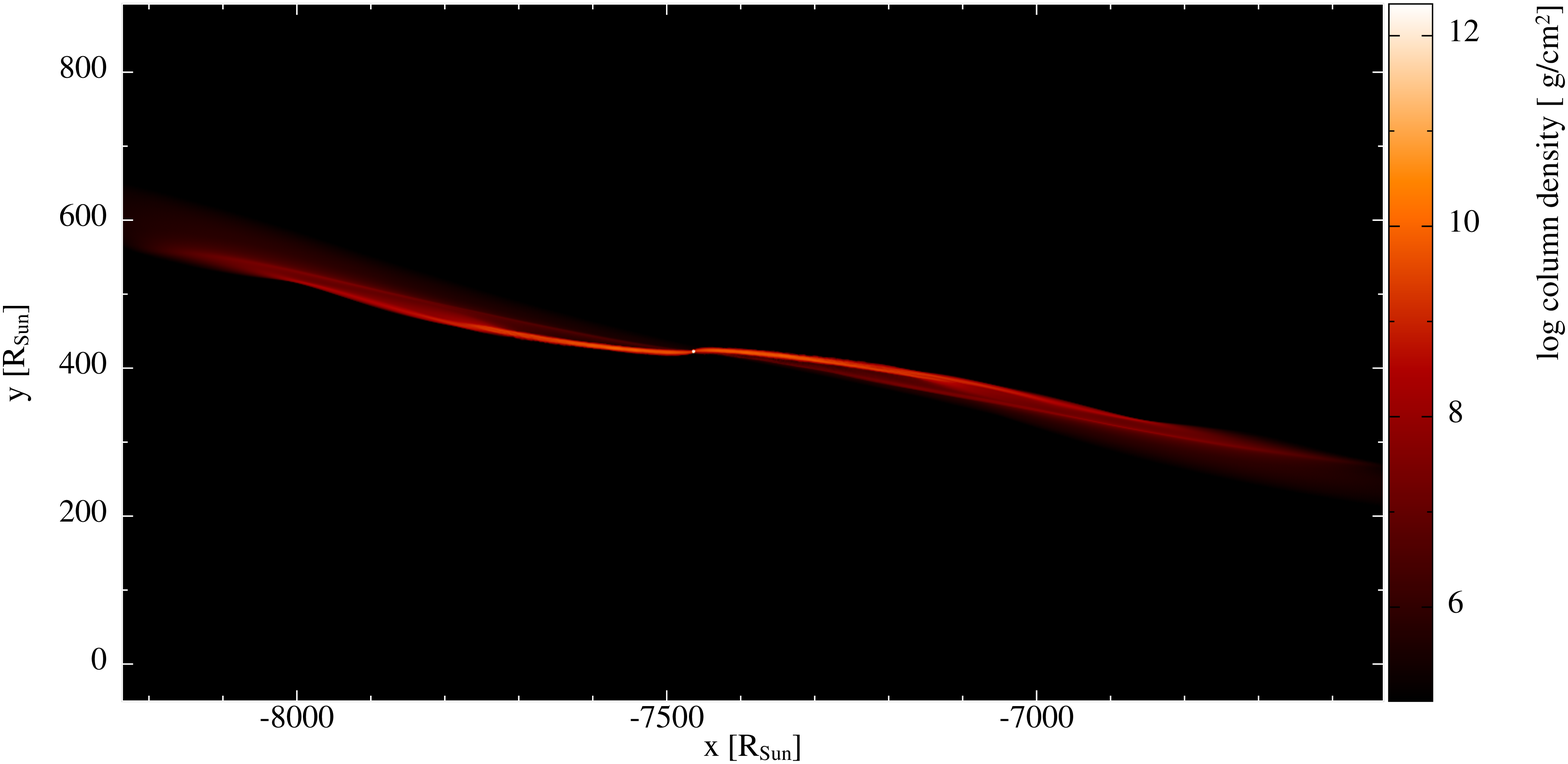}
	     \put(52.5,30){\includegraphics[width=0.16\textwidth]{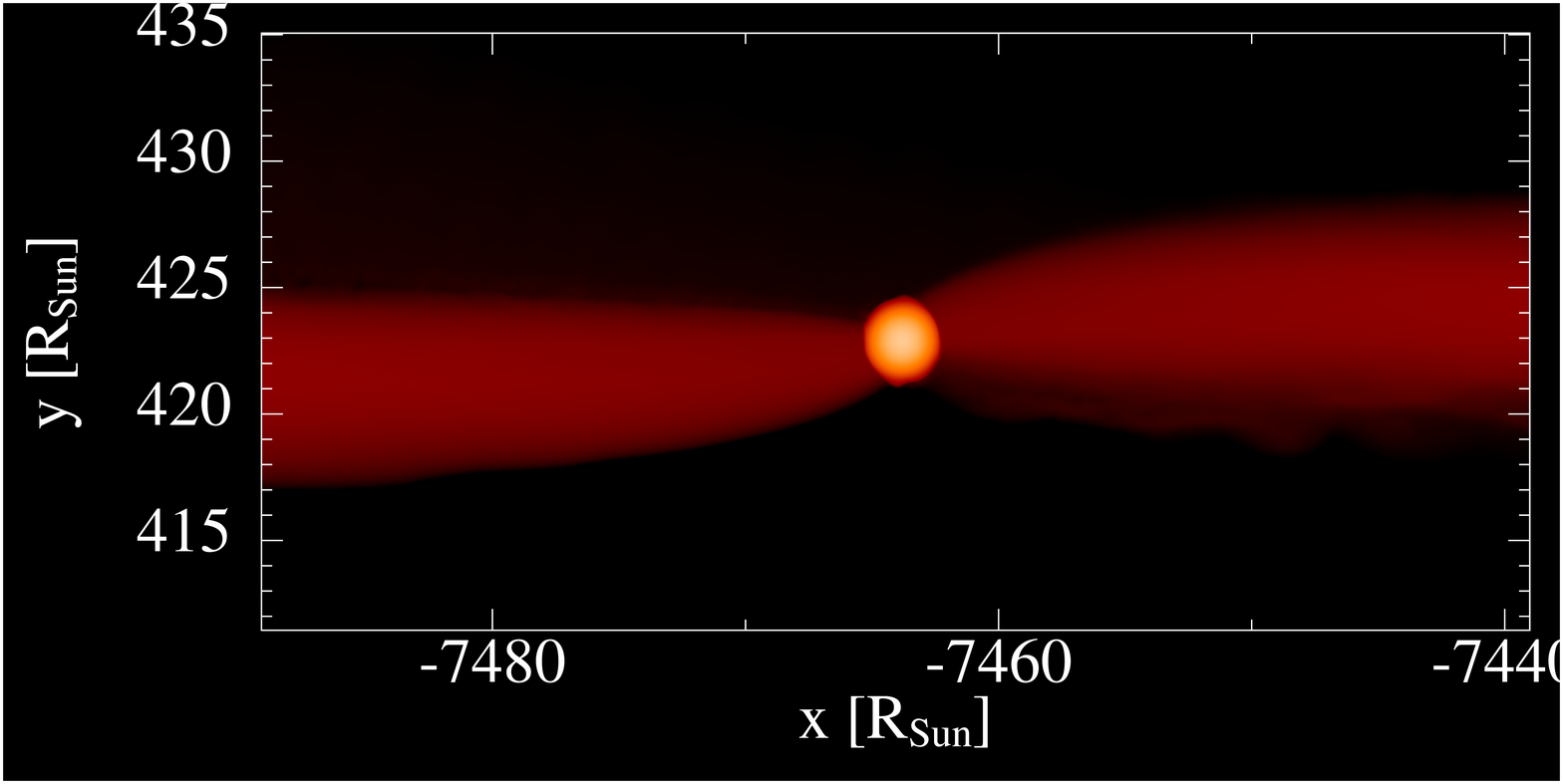}}  
  	\end{overpic}
	\caption{Debris distributions from the tidal disruption of a solar like star by a supermassive black hole with $\beta=16$. This simulation is performed with 128M particles \citep[see][for details]{Norman:2021aa}. The left panel shows the core-less debris at a time of $\sim 0.5$\,days post-pericenter, while the right panel shows the debris at a time of $\sim 5$\,days with a reformed core near the geometric centre. The inset on the right panel shows a zoom-in on the core. The core dominates the local gravity, and contains $\gtrsim 22$\% of the mass of the original star.}
	\label{fig1}
\end{figure*}

The left panel of Figure \ref{fig1} shows the disrupted debris stream at a time of $\sim 0.5$\,days post-pericenter produced by the $\beta = 16$ disruption of a solar-like, $\gamma = 5/3$ polytrope with a polytropic equation of state; this simulation used $\sim 10^8$ particles, which was the highest resolution performed (see \citealt{Norman:2021aa} for additional details of the simulation parameters). 
At this time it is clear that there is no surviving core -- the disruption appears ``complete,'' which agrees with the notion that there is a critical $\beta$ above which the star does not survive the encounter. The right panel of this figure is of the same simulation, but at a time of $\sim 5$\,days post-pericenter, and there is a single, dominant core that has reformed near the geometric center of the stream. Thus, despite being completely destroyed initially, this disruption permits the reformation of a zombie core \citep{Nixon:2021ab} under the influence of self-gravity.

\begin{figure*}
	\centering
	\includegraphics[width=0.49\textwidth]{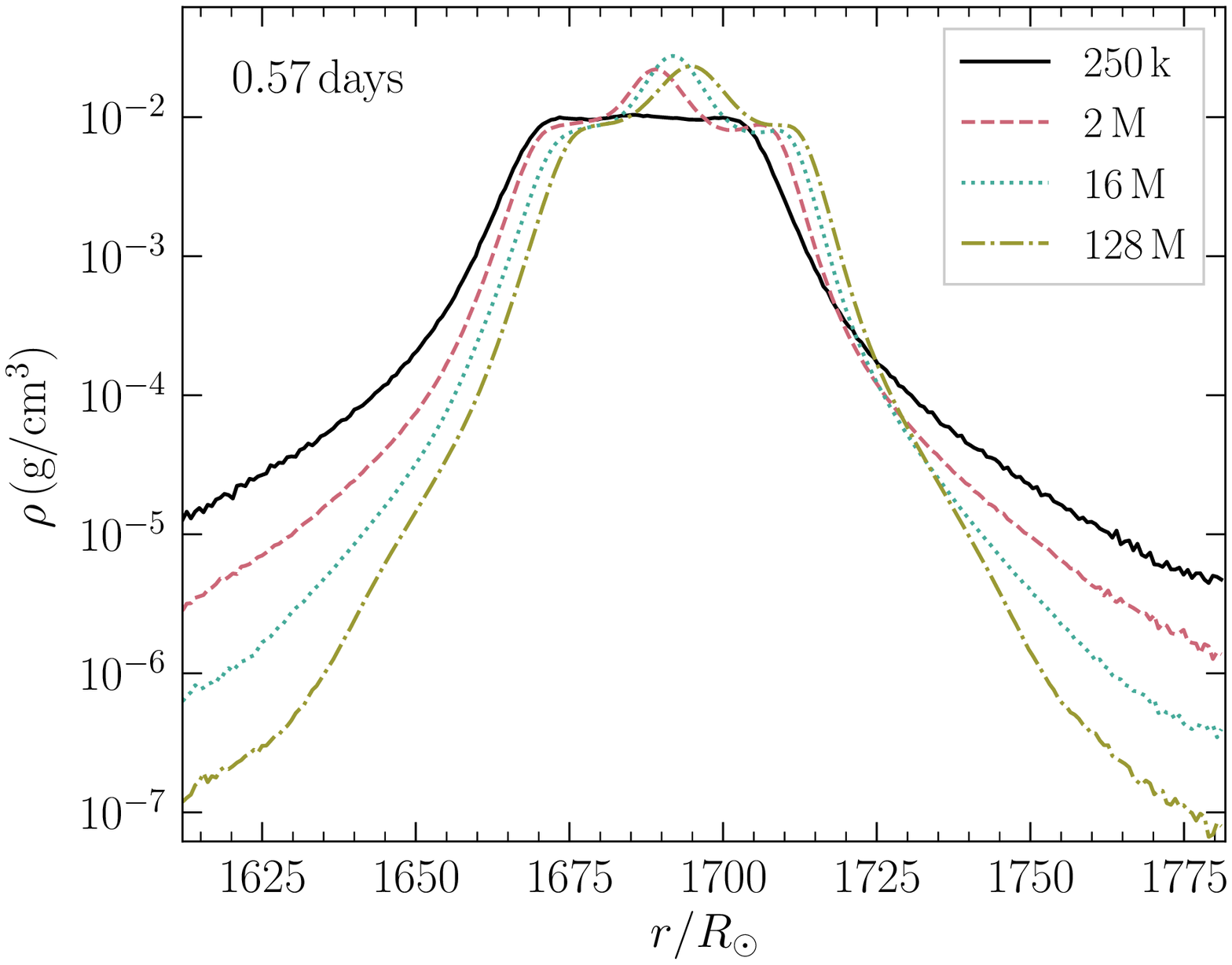}
	\includegraphics[width=0.49\textwidth]{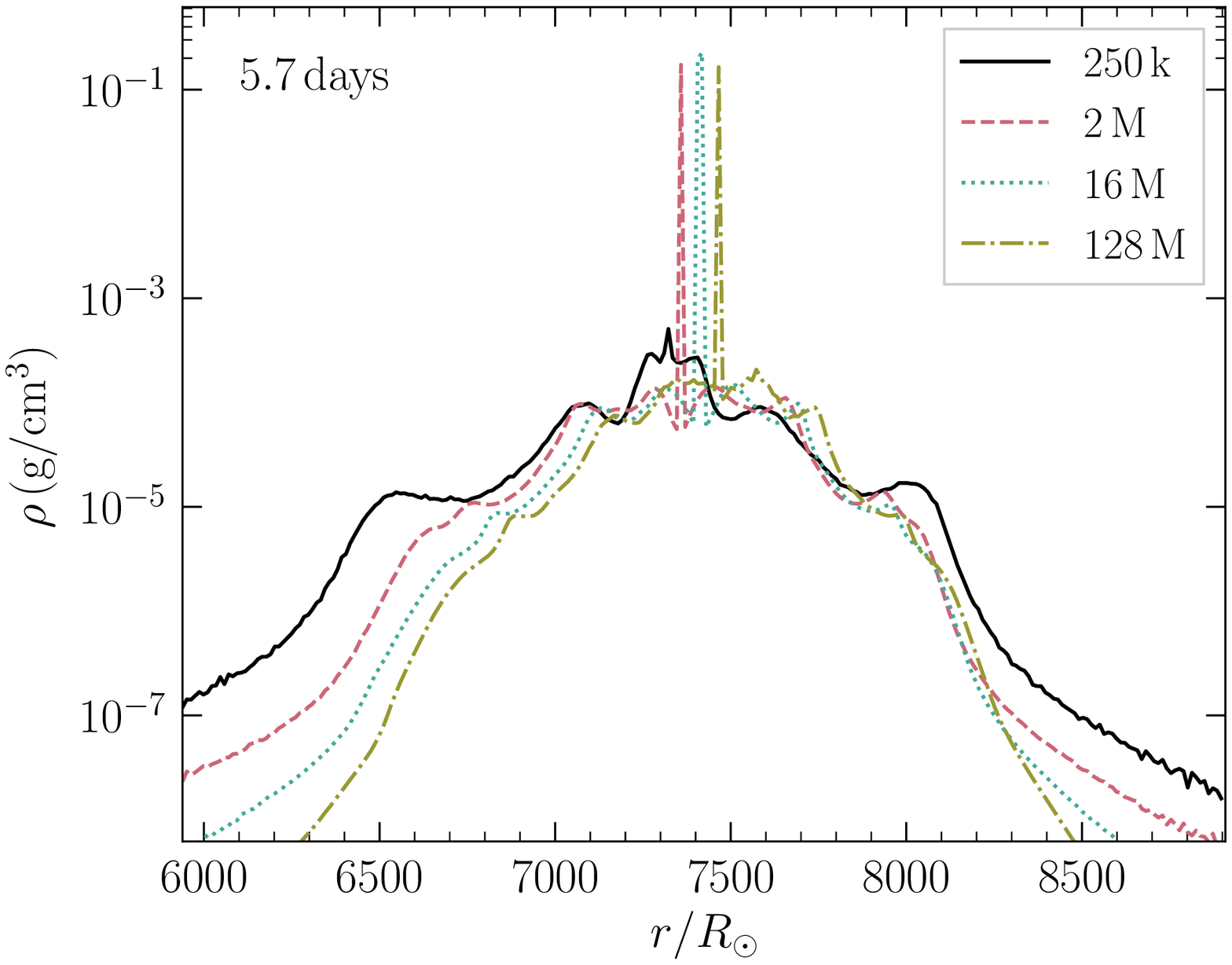}
	\caption{The density profiles of the stream corresponding to the times shown in Figure \ref{fig1}. In the left panel we can see that the star has been stretched into a long stream, of length $\sim 100\,R_\odot$ and the peak density is substantially reduced from that of the original star ($\sim 8$ g cm$^{-3}$). In the right panel we can see that a sharp peak has formed near the stream's geometric centre, corresponding to the reformed core seen in the right panel of Figure \ref{fig1}. The different curves correspond to the resolutions shown in the legend. It is clear that the properties and location of the reformed core do not depend strongly on resolution once a large enough particle number (several million) is reached.}
	\label{fig2}
\end{figure*}

Figure \ref{fig2} shows the density as a function of radial coordinate in the stream at a time of $\sim 0.5$\,days (top panel) and $\sim 5$\,days (bottom panel), where the density has been binned into 250 radial bins and averaged over its small solid angle; the different curves correspond to the resolutions shown in the legend. As a function of resolution, the location of the recollapsed core within the stream changes slightly, but neither the maximum density nor the size of the core (and, hence, its mass) changes substantially with resolution. 

\begin{figure*}
	\centering
	\includegraphics[width=0.49\textwidth]{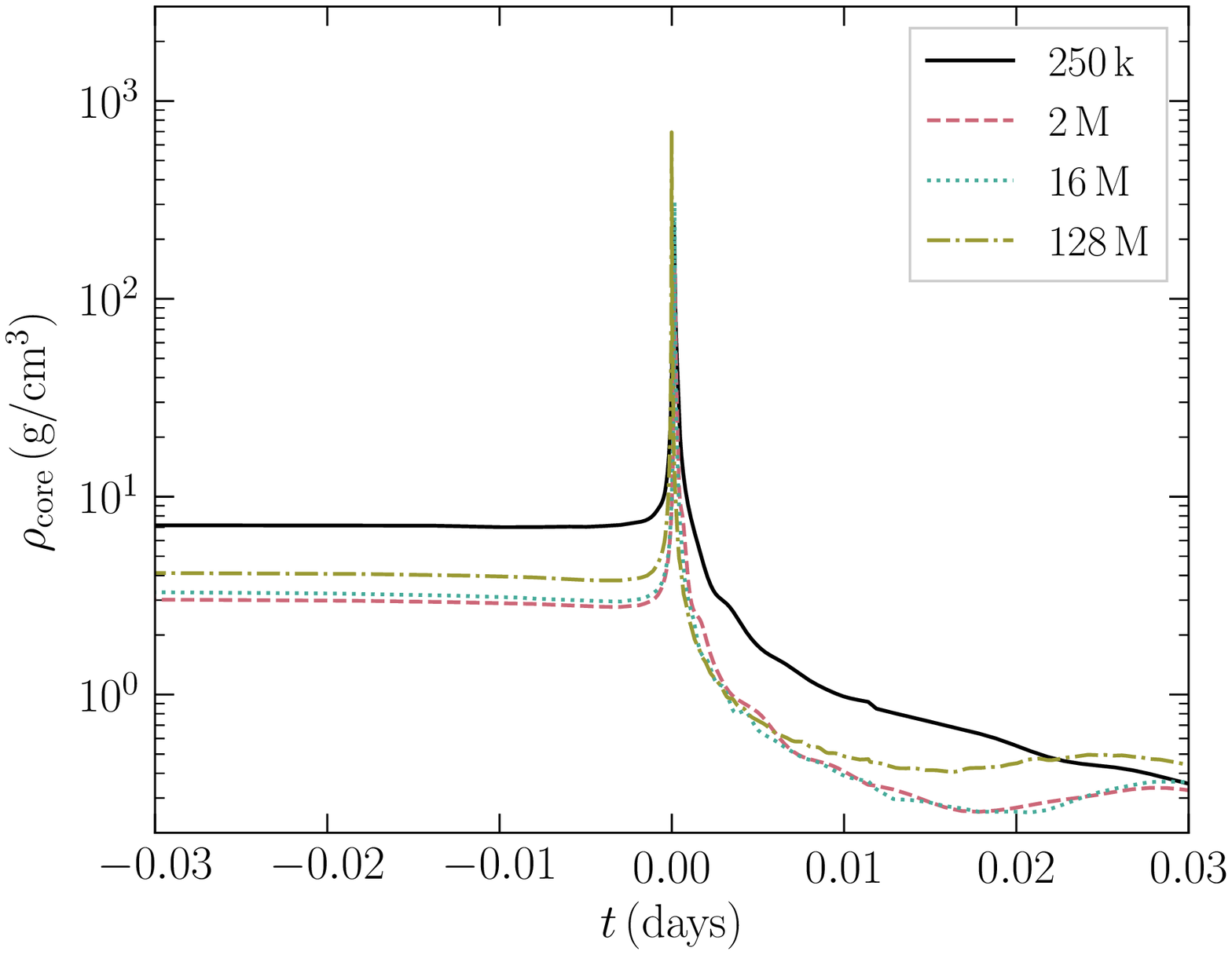}\hfill
	\includegraphics[width=0.49\textwidth]{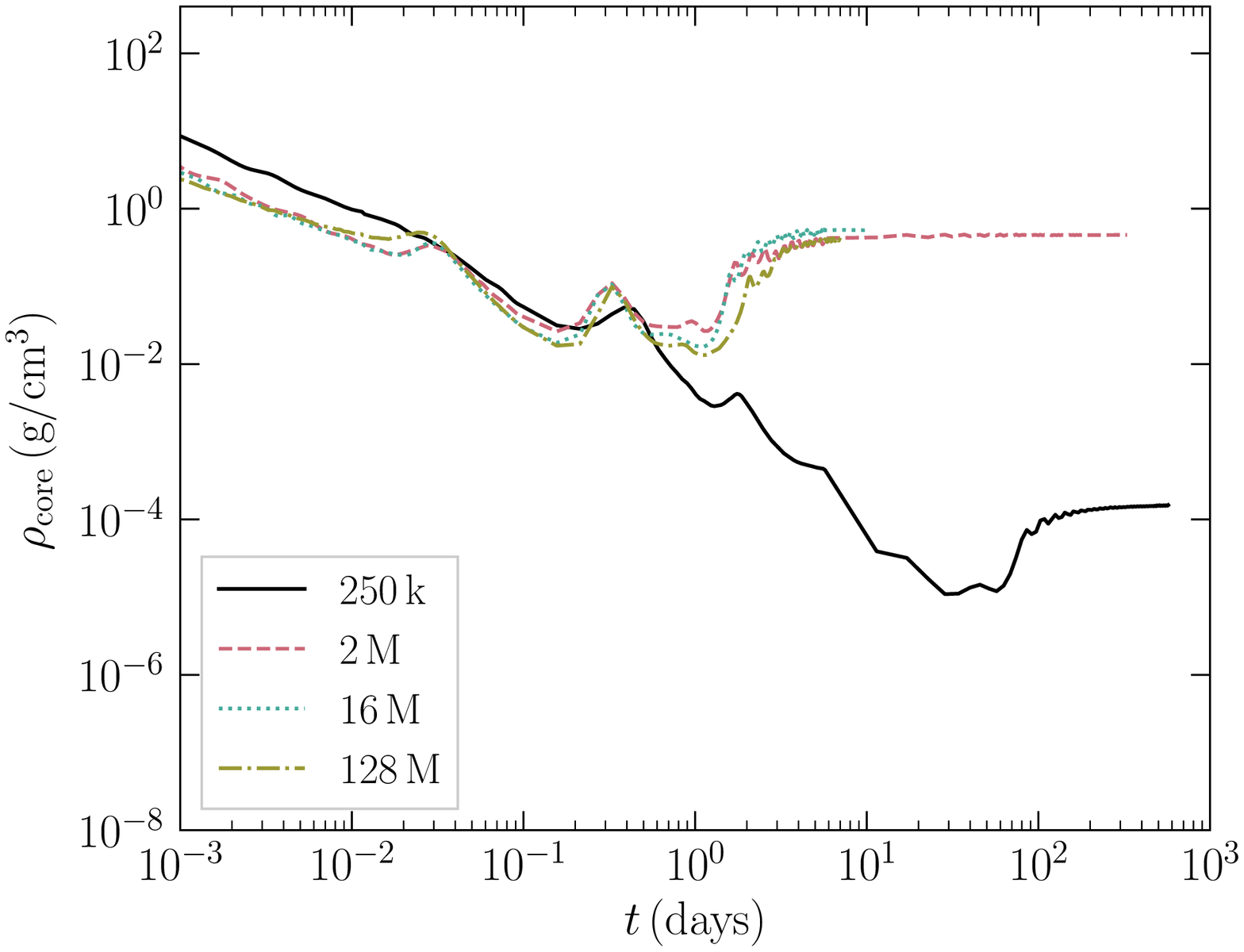}
	\caption{The temporal evolution of the density of an SPH particle that is within, and near the center of, the reformed core; the left panel is for times very near the time at which the pericenter is reached, whereas the right panel extends to later times. The spike in the left panel coincides with when the star is maximally compressed in the vertical direction, while the first relative maximum in the right panel (around a time of $\sim 0.2$ days) corresponds with the in-plane compression. The core has formed when the density plateaus near a constant value. The different curves are for the resolutions shown in the legend. }
	\label{fig3}
\end{figure*}

We postulate that the origin of the recollapse, and the revival of self-gravity, is related to the enhancement in the density that occurs as a byproduct of the vertical and in-plane compression near pericenter. 
The latter was first described in \citet{coughlin16aa}, and the former has been the subject of intense study for decades (e.g., \citealt{Carter:1983aa, Bicknell:1983aa, Stone:2013aa, coughlin21d}). In support of this suggestion, Figure \ref{fig3} shows the temporal evolution of the density of one of the SPH particles that is near the geometric center of the reformed core in the $\beta = 16$ simulation for the different resolutions; the top panel is for times very near pericenter ($t = 0$ corresponds to when the stellar center of mass reaches pericenter), while the bottom is for later times (the curves plateau to near-constant values when the core reforms). We see that there is a large spike near pericenter that signifies the vertical compression, but there is also a secondary, relative maximum soon after at a time of $\sim 0.025$ days. This time is roughly the time at which the star exits the tidal sphere on its egress, which also equals the time at which the in-plane caustic would occur in the absence of pressure (see Figure 3 of \citealt{coughlin20}). The relative maxima at later times (e.g., the one near $\sim 0.25$ days) are due to additional oscillations of the stream that are excited by these two, geometric enhancements in the density; the subsequent, runaway growth into the nonlinear regime (and the reformation of the core) is a consequence of a gravitational instability (cf.~\citealt{Coughlin:2015aa}). For further discussion of the nature of the instability and the corresponding growth rates, see \citet{coughlin20a}, and for the application to short gamma-ray bursts and the instability of the tidal tails therefrom, see \citet{coughlin20b}. 

The binding energy of the recollapsed core (to the SMBH) is negative, meaning that it will return to ($\sim$) the original pericenter of the star. However, the binding energy that we find is a function of resolution, and continues to decrease (in magnitude) as the resolution of the simulation increases. The period of the core for the lowest-resolution simulations is on the order of a couple years, while that for the highest resolution is $\sim 8$ years. Thus, while it seems likely that the reformed core will return to pericenter instead of escaping on a hyperbolic trajectory, the timescale over which it returns cannot be accurately constrained with our current simulations.

\section{Conclusions}
\label{sec:conclusions}
We analyzed the results of the late-time evolution of the $\beta = 16$ disruption of a $\gamma = 5/3$, solar-like polytrope disrupted by a $10^6M_{\odot}$ SMBH -- first presented in \citet{Norman:2021aa} -- and showed that a core reforms out of the tidally disrupted debris on a timescale of $\sim$ days post-pericenter. The fact that a core reforms at a $\beta$ that is well above the ``critical $\beta = \beta_{\rm c}$'' that separates full from partial disruption for this type of star, found to be $\beta_{\rm c} \sim 0.9$ by previous authors \citep{Guillochon:2013aa, Mainetti:2017aa, Miles:2020aa}, suggests that this tenet of TDE theory -- that self-gravity is incapable of substantially modifying the stream dynamics in the high-$\beta$ regime and that all encounters with $\beta > \beta_{\rm c}$ do not possess stellar cores -- is incorrect. 

While it was not our focus here, we also analyzed the disruption of the same star with $\beta = 8$, and while the stream displayed vigorous fragmentation into small-scale clumps (under its own self-gravity, in line with recent predictions and simulations that suggest that self-gravity is important at late times for high-$\beta$ encounters; \citealt{steinberg19, coughlin20, Nixon:2021ab}), one clump near the geometric center of the stream had a density increased above the others by a factor of $\sim 10$ and correspondingly a much larger mass than the rest. This suggests that $\beta = 8$ may also exhibit core reformation, but the mass contained in the core is insufficient to prevent further fragmentation (see the discussion in \citealt{cufari21}). More generally, this finding implies that there is nothing unique about $\beta = 16$, and that the transition between fragmentation and single core reformation, both of which are ultimately due to the same instability \citep{Coughlin:2015aa}, lies somewhere between $\beta = 8$ and 16.

A noteworthy concern is that the simulations we performed used a Newtonian gravitational field for the SMBH, while for a $10^6M_{\odot}$ SMBH, the direct capture radius (i.e., the zero-energy orbit) for a Schwarzschild black hole occurs at $4 r_{\rm g}$, where $r_{\rm g} = GM_{\bullet}/c^2$ is the gravitational radius. For a solar-like star, this gives $r_{\rm t}/r_{\rm g} \simeq 47$, so that direct capture occurs at $\beta \simeq 12$. Thus, when the SMBH mass is this large and the pericenter distance is as close to the one considered here, relativistic effects will start to play a significant role in modifying the dynamics. See the extensive discussion of general relativistic effects in \citet{Norman:2021aa, coughlin21d}. 

However, while it seems likely that relativistic effects will modify the evolution from what we have described here, it is unclear whether they will enhance or inhibit core reformation in the large-$\beta$ limit. We argued that the revival of the core is related to the increase in the density of the material as it is compressed near pericenter and post-pericenter (near $\sim r_{\rm t}$) as arises from orbital focusing. The results of \citet{Gafton:2019aa} show that the increase in the density near pericenter becomes \emph{more extreme} relative to the Newtonian approximation, with the maximum density achieved increased by a factor of $\lesssim 2$ when the stellar center of mass nears the direct capture radius. Thus, general relativistic dynamics could actually increase the prevalence of core reformation in deep TDEs, with a zombie core forming at more modest $\beta$ when general relativity is included.

On the other hand, when the orbital dynamics can be treated as non-relativistic -- which becomes increasingly accurate for large $\beta$ when the black hole mass is small -- the value of the black hole mass does not influence the compression within the tidal sphere for these types of encounters \citep{coughlin21d}. This notion suggests that core reformation is a generic feature of $\beta \simeq 16$ encounters between a $\gamma = 5/3$, polytropic star (characteristic of fully convective, low-mass stars) and a massive (compact) object. 

From the analysis in \citet{coughlin21d} and the hydrodynamical simulations in \citet{Norman:2021aa}, we conclude that any shocks formed during deep TDEs are weak and have Mach numbers $\sim 1$. The entropy generated during the stellar compression is therefore not large, and the equation of state should be physically close to adiabatic. We find that for large-$\beta$ encounters, there is significant numerical heating in the simulations, and this heating is strongly reduced as the resolution of the simulation increases (see Figure 5 of \citealt{Norman:2021aa}). Combining this result with the understanding from the analytical model in \citet{coughlin21d}, we infer that the high-$\beta$ simulations that enforce a polytropic equation of state are more representative of the (physical) encounter than those that include the entropy generated by the numerical viscosity terms. In other words, the spurious numerical heating creates a much larger pressure within and cross-sectional width of the disrupted debris stream, which lengthens the growth time of the instability and prevents the formation of a single, dominant core. However, we find that the stream in the $\beta = 16$ simulation that includes shock heating fragments into a large number of small-scale knots, indicating that the stream is still gravitationally unstable and that it is the same instability that is responsible for the core reformation and the fragmentation originally found in \citet{Coughlin:2015aa}. 

As we noted above, the core in our simulations forms on a bound orbit, but the binding energy is not converged; we therefore cannot answer when the reformed core will return to pericenter. The mass is, however, converged (or at least changes only very slightly as a function of resolution) at around $\sim 20-25$\% of the original stellar mass. The core reformation occurs sufficiently rapidly in every simulation that the process can likely be well-modeled as adiabatic, with energy losses between disruption and reformation being negligible. With the entropy unchanged (also accurate from the standpoint that the shocks that form during compression are weak; \citealt{coughlin21d,Norman:2021aa}) and an adiabatic index $\gamma = 5/3$, it follows that the density of the reformed core is

\begin{equation}
\rho_{\rm core} = \rho_{\star}\left(\frac{M_{\rm core}}{M_{\star}}\right)^{2},
\end{equation}
where $M_{\rm core}$ is the mass of the core, $M_{\star}$ is the mass of the original star, and $\rho_{\star}$ is the original stellar density. Since the tidal radius of the returning core is

\begin{equation}
r_{\rm t, core} \simeq \left(\frac{M_{\bullet}}{\rho_{\rm core}}\right)^{1/3} = r_{\rm t,\star}\left(\frac{M_{\rm core}}{M_{\star}}\right)^{-2/3},
\end{equation}
the $\beta$ of the returning-core encounter is

\begin{equation}
\beta_{\rm core} = \beta_{\star}\left(\frac{M_{\rm core}}{M_{\star}}\right)^{-2/3} \simeq 47,
\end{equation}
where in the last line we used $M_{\rm core}/M_{\star} = 0.2$. Thus, the effective $\beta$ that the surviving core experiences as it returns to pericenter is much more extreme than that of the original star. Ordinarily (i.e., according to standard wisdom) this finding would suggest that the reformed core is completely destroyed by tides upon its pericenter passage. Our present results, however, call this conclusion into question; it could instead be the case that this even more extreme $\beta$ results in yet another reformed core post-pericenter (i.e., if $\beta = 16$ results in a reformed core, then it may well be the case that $\beta = 47$ does as well), with each successive disruption producing a reformed core with progressively less mass and having a correspondingly even more extreme $\beta$. The recurrent disruptions at each pericenter passage could lead to periodic flaring events such as those documented in the recent literature \citep{miniutti19, song20, arcodia21, chakraborty21, payne21a, payne21b}.

\acknowledgements
CJN acknowledges funding from the European Union’s Horizon 2020 research and innovation program under the Marie Sk\l{}odowska-Curie grant agreement No 823823 (Dustbusters RISE project). E.R.C. acknowledges support from the National Science Foundation through grant AST-2006684. This research used the ALICE High Performance Computing Facility at the University of Leicester. This work was performed using the DiRAC Data Intensive service at Leicester, operated by the University of Leicester IT Services, which forms part of the STFC DiRAC HPC Facility (\url{www.dirac.ac.uk}). The equipment was funded by BEIS capital funding via STFC capital grants ST/K000373/1 and ST/R002363/1 and STFC DiRAC Operations grant ST/R001014/1. DiRAC is part of the National e-Infrastructure. We used {\sc splash} \citep{Price:2007aa} for some of the figures.


\bibliographystyle{aasjournal}
\bibliography{nixon}

\begin{thebibliography}{}
\expandafter\ifx\csname natexlab\endcsname\relax\def\natexlab#1{#1}\fi
\providecommand{\url}[1]{\href{#1}{#1}}

\bibitem[{{Arcodia} {et~al.}(2021){Arcodia}, {Merloni}, {Nandra}, {Buchner},
  {Salvato}, {Pasham}, {Remillard}, {Comparat}, {Lamer}, {Ponti}, {Malyali},
  {Wolf}, {Arzoumanian}, {Bogensberger}, {Buckley}, {Gendreau}, {Gromadzki},
  {Kara}, {Krumpe}, {Markwardt}, {Ramos-Ceja}, {Rau}, {Schramm}, \&
  {Schwope}}]{arcodia21}
{Arcodia}, R., {Merloni}, A., {Nandra}, K., {et~al.} 2021, \nat, 592, 704

\bibitem[{{Bicknell} \& {Gingold}(1983)}]{Bicknell:1983aa}
{Bicknell}, G.~V., \& {Gingold}, R.~A. 1983, \apj, 273, 749

\bibitem[{{Carter} \& {Luminet}(1983)}]{Carter:1983aa}
{Carter}, B., \& {Luminet}, J.~P. 1983, \aap, 121, 97

\bibitem[{{Chakraborty} {et~al.}(2021){Chakraborty}, {Kara}, {Masterson},
  {Giustini}, {Miniutti}, \& {Saxton}}]{chakraborty21}
{Chakraborty}, J., {Kara}, E., {Masterson}, M., {et~al.} 2021, \apjl, 921, L40

\bibitem[{{Coughlin} \& {Nixon}(2015)}]{Coughlin:2015aa}
{Coughlin}, E.~R., \& {Nixon}, C. 2015, \apjl, 808, L11

\bibitem[{{Coughlin} \& {Nixon}(2021)}]{coughlin21d}
---. 2021, ApJ~in~press, arXiv:2111.12736

\bibitem[{{Coughlin} {et~al.}(2016){Coughlin}, {Nixon}, {Begelman}, {Armitage},
  \& {Price}}]{coughlin16aa}
{Coughlin}, E.~R., {Nixon}, C., {Begelman}, M.~C., {Armitage}, P.~J., \&
  {Price}, D.~J. 2016, \mnras, 455, 3612

\bibitem[{{Coughlin} \& {Nixon}(2020)}]{coughlin20a}
{Coughlin}, E.~R., \& {Nixon}, C.~J. 2020, \apjs, 247, 51

\bibitem[{{Coughlin} {et~al.}(2020{\natexlab{a}}){Coughlin}, {Nixon}, {Barnes},
  {Metzger}, \& {Margutti}}]{coughlin20b}
{Coughlin}, E.~R., {Nixon}, C.~J., {Barnes}, J., {Metzger}, B.~D., \&
  {Margutti}, R. 2020{\natexlab{a}}, \apjl, 896, L38

\bibitem[{{Coughlin} {et~al.}(2020{\natexlab{b}}){Coughlin}, {Nixon}, \&
  {Miles}}]{coughlin20}
{Coughlin}, E.~R., {Nixon}, C.~J., \& {Miles}, P.~R. 2020{\natexlab{b}}, \apjl,
  900, L39

\bibitem[{{Cufari} {et~al.}(2022){Cufari}, {Coughlin}, \& {Nixon}}]{cufari21}
{Cufari}, M., {Coughlin}, E.~R., \& {Nixon}, C.~J. 2022, \apj, 924, 34

\bibitem[{{Fabian} {et~al.}(1975){Fabian}, {Pringle}, \& {Rees}}]{fabian75}
{Fabian}, A.~C., {Pringle}, J.~E., \& {Rees}, M.~J. 1975, \mnras, 172, 15

\bibitem[{{Gafton} \& {Rosswog}(2019)}]{Gafton:2019aa}
{Gafton}, E., \& {Rosswog}, S. 2019, \mnras, 487, 4790

\bibitem[{{Guillochon} \& {Ramirez-Ruiz}(2013)}]{Guillochon:2013aa}
{Guillochon}, J., \& {Ramirez-Ruiz}, E. 2013, \apj, 767, 25

\bibitem[{{Lee} \& {Ostriker}(1986)}]{lee86}
{Lee}, H.~M., \& {Ostriker}, J.~P. 1986, \apj, 310, 176

\bibitem[{{Mainetti} {et~al.}(2017){Mainetti}, {Lupi}, {Campana}, {Colpi},
  {Coughlin}, {Guillochon}, \& {Ramirez-Ruiz}}]{Mainetti:2017aa}
{Mainetti}, D., {Lupi}, A., {Campana}, S., {et~al.} 2017, \aap, 600, A124

\bibitem[{{Miles} {et~al.}(2020){Miles}, {Coughlin}, \& {Nixon}}]{Miles:2020aa}
{Miles}, P.~R., {Coughlin}, E.~R., \& {Nixon}, C.~J. 2020, \apj, 899, 36

\bibitem[{{Miniutti} {et~al.}(2019){Miniutti}, {Saxton}, {Giustini},
  {Alexander}, {Fender}, {Heywood}, {Monageng}, {Coriat}, {Tzioumis}, {Read},
  {Knigge}, {Gandhi}, {Pretorius}, \& {Ag{\'\i}s-Gonz{\'a}lez}}]{miniutti19}
{Miniutti}, G., {Saxton}, R.~D., {Giustini}, M., {et~al.} 2019, \nat, 573, 381

\bibitem[{{Nixon} {et~al.}(2021){Nixon}, {Coughlin}, \& {Miles}}]{Nixon:2021ab}
{Nixon}, C.~J., {Coughlin}, E.~R., \& {Miles}, P.~R. 2021, \apj, In Press

\bibitem[{{Norman} {et~al.}(2021){Norman}, {Nixon}, \&
  {Coughlin}}]{Norman:2021aa}
{Norman}, S.~M.~J., {Nixon}, C.~J., \& {Coughlin}, E.~R. 2021, \apj, 923, 184

\bibitem[{{Ogilvie}(2014)}]{ogilvie14}
{Ogilvie}, G.~I. 2014, \araa, 52, 171

\bibitem[{{Payne} {et~al.}(2021{\natexlab{a}}){Payne}, {Shappee}, {Hinkle},
  {Vallely}, {Kochanek}, {Holoien}, {Auchettl}, {Stanek}, {Thompson},
  {Neustadt}, {Tucker}, {Armstrong}, {Brimacombe}, {Cacella}, {Cornect},
  {Denneau}, {Fausnaugh}, {Flewelling}, {Grupe}, {Heinze}, {Lopez}, {Monard},
  {Prieto}, {Schneider}, {Sheppard}, {Tonry}, \& {Weiland}}]{payne21a}
{Payne}, A.~V., {Shappee}, B.~J., {Hinkle}, J.~T., {et~al.} 2021{\natexlab{a}},
  \apj, 910, 125

\bibitem[{{Payne} {et~al.}(2021{\natexlab{b}}){Payne}, {Shappee}, {Hinkle},
  {Holoien}, {Auchettl}, {Kochanek}, {Stanek}, {Thompson}, {Tucker},
  {Armstrong}, {Boyd}, {Brimacombe}, {Cornect}, {Huber}, {Jha}, \&
  {Lin}}]{payne21b}
---. 2021{\natexlab{b}}, arXiv e-prints, arXiv:2104.06414

\bibitem[{{Press} \& {Teukolsky}(1977)}]{press77}
{Press}, W.~H., \& {Teukolsky}, S.~A. 1977, \apj, 213, 183

\bibitem[{{Price}(2007)}]{Price:2007aa}
{Price}, D.~J. 2007, \pasa, 24, 159

\bibitem[{{Song} {et~al.}(2020){Song}, {Shu}, {Sun}, {Xue}, {Jin}, {Zhang},
  {Jiang}, {Dou}, \& {Wang}}]{song20}
{Song}, J.~R., {Shu}, X.~W., {Sun}, L.~M., {et~al.} 2020, \aap, 644, L9

\bibitem[{{Steinberg} {et~al.}(2019){Steinberg}, {Coughlin}, {Stone}, \&
  {Metzger}}]{steinberg19}
{Steinberg}, E., {Coughlin}, E.~R., {Stone}, N.~C., \& {Metzger}, B.~D. 2019,
  \mnras, 485, L146

\bibitem[{{Stone} {et~al.}(2013){Stone}, {Sari}, \& {Loeb}}]{Stone:2013aa}
{Stone}, N., {Sari}, R., \& {Loeb}, A. 2013, \mnras, 435, 1809

\end{thebibliography}

\end{document}